\documentclass[%
 reprint,
 amsmath,amssymb,
 aps,
]{revtex4-1}

\usepackage{graphicx}
\usepackage{dcolumn}
\usepackage{bm}

 \usepackage{fullpage}
\usepackage{float}
  \usepackage{latexsym}
  \usepackage{epsf}
   \usepackage{amsmath,amssymb,amsthm}

\usepackage[swedish, english]{babel}
\usepackage[utf8]{inputenc}
  \usepackage{tikz}

\usepackage{enumerate}
  \usepackage{verbatim}
  \usepackage{stmaryrd}
\usepackage{mathtools}
\usepackage{tensor}
\usepackage[all]{xy}

\usepackage[colorlinks=true
,urlcolor=blue
,anchorcolor=black
,citecolor=black
,filecolor=black
,linkcolor=black
,menucolor=black
,linktocpage=true
,pdfproducer=medialab
,bookmarks=false]{hyperref}

\DeclareMathOperator{\bk}{\mathbf{k}}
\DeclareMathOperator{\bB}{\mathbf{B}}
\DeclareMathOperator{\bA}{\mathbf{A}}

\usepackage[left=1.2cm, top=2cm, right=1.2cm, textwidth=10cm]{geometry}
\usepackage{physics}
\usepackage{hyperref}
\begin{document}


\title{Magneto-optical conductivity in generic Weyl semimetals}

\author{Marcus St\aa lhammar$^1$}
 \email{marcus.stalhammar@fysik.su.se}
\author{Jorge Larana-Aragon$^1$}
\author{Johannes Knolle$^{2,3,4}$}
\author{Emil J. Bergholtz$^1$}
\affiliation{$^1$Department of Physics, Stockholm University, AlbaNova University Center, 106 91 Stockholm, Sweden}
\affiliation{$^{2}$Department of Physics TQM, Technische Universit\"at M\"unchen, James-Franck-Straße 1, D-85748 Garching, Germany}
\affiliation{$^{3}$Munich Center for Quantum Science and Technology (MCQST), 80799 Munich, Germany}
\affiliation{$^{4}$Blackett Laboratory, Imperial College London, London SW7 2AZ, United Kingdom}


\begin{abstract}
Magneto-optical studies of Weyl semimetals have been proposed as a versatile tool for observing low-energy Weyl fermions in  candidate materials including the chiral Landau level. However, previous theoretical results have been restricted to the linearized regime around the Weyl node and are at odds with experimental findings. Here, we derive a closed form expression for the magneto-optical conductivity of generic Weyl semimetals in the presence of an external magnetic field aligned with the tilt of the spectrum. The systems are taken to have linear dispersion in two directions, while the tilting direction can consist of any arbitrary continuously differentiable function. This general calculation is then used to analytically evaluate the magneto-optical conductivity of Weyl semimetals expanded to cubic order in momentum. In particular, systems with arbitrary tilt, as well as systems hosting trivial Fermi pockets are investigated. The higher-order terms in momentum close the Fermi pockets in the type-II regime, removing the need for unphysical cutoffs when evaluating the magneto-optical conductivity. Crucially, the ability to take into account closed over-tilted and  additional trivial Fermi pockets  allows us to treat model systems closer to actual materials and we propose a simple explanation why the presence of parasitic trivial Fermi pockets can mask the characteristic signature of Weyl fermions in magneto-optical conductivity measurements.

\end{abstract}

\maketitle

\section{Introduction}\label{sec:I}

Weyl fermions, charge carrying massless particles, were theoretically predicted in 1929 by Hermann Weyl in the context of particle physics \cite{W1929}. These elusive particles remained a theoretical concept for nearly a century before they were experimentally realized as quasi particles in condensed matter systems naturally called Weyl semimetals \cite{XBAN2015,LWFM2015,LWYRFJS2015}. They appear as nondegenerate intersections between the valence and the conduction band and come as two different types \cite{AMV2018}: type-I which is an actual semimetallic phase with point like Fermi surface \cite{WTVS2011,HQ2013,HZLW2015}  and type-II whose dispersion is over-tilted, resulting in Fermi pockets in connection with the band intersection \cite{BLTMU2015,SGWWTDB2015}. These materials were predicted to have intriguing transport properties, ranging from the chiral anomaly \cite{HZLW2015,ZXBY2016,NM1983,XPYQ2015,SY2012,G2012,ZB2012,GT2013,PGAPV2014,KGM2015,KGM2017,BKS2018} to nonsaturated magnetoresistence and conductivity  \cite{AXFT2014,KSXMY2017,Zetal2016}. 

Early on it was suggested that the magneto-optical conductivity could serve as an experimental tool for observing salient features of Weyl semimetals. For example, the special Landau level structure leads to a series of asymmetric sharp peaks on top of a linear background and transitions from the chiral level result in a characteristic hump at low frequencies \cite{AC2013}. Unfortunately, these crisp theoretical predictions of transport properties are difficult to reconcile with the experimental measurements~\cite{PGW2020,XZC2018} and the characteristic peaks from sharp Landau level transitions are generically not found, see Ref.~\onlinecite{PD2020} for a recent review. To address this point, we investigate the effects of higher-order terms in momenta in the Hamiltonian allowing us to study more generic Weyl model systems. 

Our study goes beyond previous theoretical works where mainly systems linear in momenta were considered \cite{AC2013,TCG2016,UB2016,YYY2016}. Here, we include higher order corrections, which for example make the Fermi pockets in type-II Weyl semimetals finite in size which avoids an unphysical cut-off in momentum. We also show that our approach can treat systems with trivial Fermi pockets, i.e., Fermi pockets not occurring in direct contact with the Weyl node as is often the case in experiments~\cite{PD2020}. To this end, we derive an analytical closed-form expression for the magneto-optical conductivity, which can be applied to a large family of Weyl-like systems. We apply our general framework to a number of representative examples, i.e.,  untilted systems, tilted type-I and type-II systems, and systems hosting trivial Fermi pockets---all which are expanded to cubic order in momentum. The latter two represent inequivalent ways of over-tilting the Weyl cone, and we investigate how the magneto-optical conductivity is affected. Finally, we propose a simple picture why the presence of trivial pockets can mask the characteristic response of the Weyl fermions.

The rest of the paper is structured as follows. In Sec. \ref{sec:II} the general setup is described, taking the form of a generic noninteracting two band model in three dimensions, whose dispersion is conical in two directions and consists of arbitrary $C^1$-functions in the tilt direction. Also, the analytical calculation of the magneto-optical conductivity for this class of systems in a magnetic field aligned with the tilt is presented. Then, in Sec. \ref{sec:III}, this expression is evaluated for systems with cubic order corrections in the direction of the tilt and magnetic field. Both untilted systems, tilted type-I and type-II systems as well as systems hosting trivial Fermi pockets are examined. In Sec. \ref{sec:IV} we interpret the results, discuss the experimental relevance and suggest future directions.

\section{General Setup}\label{sec:II}

We begin by describing the basic ingredients for our theoretical description of Weyl semimetals. Initially, we consider a generic Weyl-like system described by the Hamiltonian
\begin{equation} \label{eq:genham}
    H(\bk) = \hbar v_F\left[ k_x\sigma_x + k_y \sigma_y + g(k_z) \sigma_z + h(k_z) \sigma_0\right],
\end{equation}
where $\sigma_i$ for $i=x,y,z$ are the Pauli matrices, $\sigma_0$ is the $2\times 2$ identity matrix, $g(k_z)$ and $h(k_z)$ are any continuously differentiable functions of $k_z$ and $v_F$ is the Fermi velocity. The tilt of the dispersion is given by $h'(k_{\text{W}})$, where $k_{\text{W}}$ denotes the corresponding Weyl node.

For concreteness, we consider the effect of a magnetic field oriented along the direction of $k_z$, $\bB = (0,0,B)$. In the Landau gauge, the corresponding vector potential reads $\bA = (0,Bx,0)$, which is introduced into the Hamiltonian as $\Pi_i=\hbar k_i-\frac{e}{c}A_i$. By introducing the standard creation and annihilation operators $a=\frac{l_B}{\sqrt{2}\hbar}\left(\Pi_x-i\Pi_y\right)$ and $a^{\dagger}=\frac{l_B}{\sqrt{2}\hbar}\left(\Pi_x+i\Pi_y\right)$, with $[a,a^{\dagger}]=1$ and defining $l_B:=\sqrt{\frac{\hbar}{eB}}$, the effective Hamiltonian generically reads
\begin{equation} \label{eq:hambfield}
    H=\hbar v_F\begin{pmatrix}h(k_z)+g(k_z) & \frac{\sqrt{2}}{l_B}a^{\dagger}\\\frac{\sqrt{2}}{l_B}a & h(k_z)-g(k_z) \end{pmatrix},
\end{equation}
which has the following eigenvalues:
\begin{equation}
    E_{n,\lambda}(k_z,l_B) = \hbar v_F\left[ h(k_z)+\lambda\sqrt{g^2(k_z)+\frac{2n}{l^2_B}}\right],
\end{equation}
where $n$ denotes the Landau level and $\lambda=\pm1$. The former expression is only valid for $n\neq0$. For $n=0$, the chiral energy level reads
\begin{equation}
    E_{0}(k_z) = \hbar v_F\left(h(k_z)+g(k_z)\right).
\end{equation}
The eigenstates have the form $\psi_{n,\lambda}(k_z)=\begin{pmatrix}\lambda u_{n,\lambda}(k_z) \\ v_{n,\lambda}(k_z)\end{pmatrix}$ where
\begin{align}
    u_{n,\lambda}(k_z) &= \sqrt{\frac{1}{2}\left[1+\frac{g(k_z)}{\lambda \sqrt{g^2(k_z)+\frac{2n}{l_B^2}}}\right]}, \label{eq:es1}
    \\
    v_{n,\lambda}(k_z) &= \sqrt{\frac{1}{2}\left[1-\frac{g(k_z)}{\lambda \sqrt{g^2(k_z)+\frac{2n}{l_B^2}}}\right]}, \label{eq:es2}
\end{align}
for $n\neq 0$, and 
\begin{equation} \label{eq:es3}
    \psi_0=\begin{pmatrix} 1\\0\end{pmatrix},
\end{equation}
for $n=0$.

\subsection{Response Function}
In the one-loop approximation, the magneto-optical conductivity is obtained from the linear response function which reads,
\begin{align}\label{eq:response function}
    &\chi_{\alpha\beta}(\omega)  \nonumber
    \\
    &=\frac{1}{2\pi l^2_B}\sum_{n,n'}\sum_{\lambda,\lambda '}\int \frac{dk_z}{2\pi}\frac{f\left[E_{n,\lambda}(k_z)\right]-f\left[E_{n',\lambda '}(k_z)\right]}{\hbar\omega+E_{n,\lambda}(k_z)-E_{n',\lambda '}(k_z)+i\epsilon}  \nonumber
    \\
    & \times\langle{\psi_{n,\lambda}(k_z)}|j_{\alpha}|{\psi_{n',\lambda '}(k_z)}\rangle\langle{\psi_{n',\lambda'}(k_z)}|j_{\beta}|{\psi_{n,\lambda }(k_z)}\rangle,
\end{align}
where $f(E)=\frac{1}{1+e^{\frac{E-\mu}{k_B T}}}$ is the Fermi-Dirac distribution function, $T$ is the temperature and $j_{\alpha}$ is a current operator, given by 
\begin{equation}
    j_{\alpha}=\frac{e}{\hbar}\frac{\partial H}{\partial \Pi_{\alpha}}.
\end{equation}

We will focus on the transverse components of the response function, i.e., $\chi_{xx}$ and $\chi_{xy}$. Due to rotational symmetry, the components $\chi_{yy}=\chi_{xx}$ and $\chi_{yx}=\chi_{xy}$. The matrix elements in Eq. \eqref{eq:response function} are given by
\begin{align}
    &\quad\langle{\psi_{n,\lambda}(k_z)}|j_{x}|{\psi_{n',\lambda '}(k_z)}\rangle\langle{\psi_{n',\lambda'}(k_z)}|j_{x}|{\psi_{n,\lambda }(k_z)}\rangle
    \nonumber
    \\
    &=e^2v_F^2\left\{\left[u_{n,\lambda}(k_z)\right]^2\left[v_{n',\lambda'}(k_z)\right]^2\delta_{n+1,n'}\right. \nonumber 
    \\
    &\quad +\left. \left[v_{n,\lambda}(k_z)\right]^2\left[u_{n',\lambda'}(k_z)\right]^2\delta_{n-1,n'}\right\},
    \end{align}
    
    \begin{align}
    &\quad\langle{\psi_{n,\lambda}(k_z)}|j_{x}|{\psi_{n',\lambda '}(k_z)}\rangle\langle{\psi_{n',\lambda'}(k_z)}|j_{y}|{\psi_{n,\lambda }(k_z)}\rangle
    \nonumber
    \\
    &=ie^2v_F^2\left\{\left[u_{n,\lambda}(k_z)\right]^2\left[v_{n',\lambda'}(k_z)\right]^2\delta_{n+1,n'}  \nonumber \right.
    \\
    &\quad-\left. \left[v_{n,\lambda}(k_z)\right]^2\left[u_{n',\lambda'}(k_z)\right]^2\delta_{n-1,n'}\right\}.
\end{align}
Note, the structure of the matrix elements directly dictates the allowed magneto-optical transitions---the only allowed Landau level transitions are $n\to n\pm 1$ between the same chirality ($\lambda=\lambda'$ intraband) or between different chiralities ($\lambda\neq\lambda'$ interband).

To continue using analytical methods, we proceed in the following way. First, we invoke the Kramers-Kronig relations, which relates the real and imaginary parts of the response function. Thus, the full information of the response function is captured in both its real part and imaginary part independently. Hence, it is sufficient to compute e.g. the imaginary part of $\chi_{xx}$ and the real part of $\chi_{xy}$. Second, we will consider the clean limit, $\epsilon \to 0$, allowing us to perform the integration over $k_z$ and providing an upper bound on the Landau level summation. Detailed explanations and calculations are found in the supplementary material, Appendices \ref{app:I} and \ref{app:II}. After some tedious, yet straight forward calculations, we finally arrive at
\begin{widetext}
\begin{align} \label{eq:resfunxx}
\text{Im}\left[\chi_{xx}(\omega)\right]&=\frac{e^2v_F^2}{16\pi l^2_B}\sum_{n=0}^{\left\lfloor \frac{\left(2 v_F^2-\omega^2l^2_B \right)^2}{8v_F^2\omega^2l^2_B}\right\rfloor}\sum_{i=1}^{2m}\left(\left[D^+_n(k_i,\omega,l_B)+D^-_n(k_i,\omega,l_B)\right]C_{n,n+1}(k_i)\right. 
\nonumber
\\
&\quad \times \left.\left\{\left[F^{-+}_{n,n+1}(k_i)+F^{+-}_{n,n+1}(k_i)\right]\theta(\frac{2v_F^2}{l^2_B}-\omega^2)+\left[F^{++}_{n,n+1}(k_i)+F^{--}_{n,n+1}(k_i)\right]\theta(\omega^2-\frac{2v_F^2}{l^2_B})\right\}\right),
\end{align}
\begin{align} \label{eq:resfunxy}
    \text{Re}\left[\chi_{xy}(\omega)\right]&= -\frac{e^2v_F^2}{16\pi l^2_B}\sum_{n=0}^{\left\lfloor \frac{\left(2 v_F^2-\omega^2l^2_B \right)^2}{8v_F^2\omega^2l^2_B}\right\rfloor}\sum_{i=1}^{2m}\left(\left[D^+_n(k_i,\omega,l_B)-D^-_n(k_i,\omega,l_B)\right]C_{n,n+1}(k_i) \right.\times
    \nonumber
    \\
    &\quad \left.\left\{\left[F^{-+}_{n,n+1}(k_i)+F^{+-}_{n,n+1}(k_i)\right]\theta(\frac{2v_F^2}{l^2_B}-\omega^2)+\left[F^{++}_{n,n+1}(k_i)+F^{--}_{n,n+1}(k_i)\right]\theta(\omega^2-\frac{2v_F^2}{l^2_B})\right\}\right),
\end{align}
where we have defined the following dimensionless coefficients:
\begin{align}
    C_{n,n+1}&=C_{n+1,n}=\abs{\frac{\left(\frac{\omega^2}{4v_{F}^{2}}-\frac{v_{F}^{2}}{\omega^2l_{B}^4}\right)}{g(k_i)g'(k_i)\frac{\omega}{v_F}}},
    \\
    D^{\pm} &= \frac{\sinh{\frac{\hbar\omega}{2k_BT}}}{\cosh{\left[\frac{2\hbar v_F h(k_i)\pm\frac{2\hbar^2v_F^2}{\hbar\omega l^2_B}-2\mu}{2k_BT}\right]}+\cosh\frac{\hbar\omega}{2k_BT}},
    \\
    F^{\pm\pm}_{n,m} &= \left(1\pm \frac{\tilde{g}(k_i)}{\omega\sqrt{g^2(k_i)+\frac{2n}{l^2_B}}}\right)\left(1\pm\frac{\tilde{g}(k_i)}{\omega\sqrt{g^2(k_i)+\frac{2m}{l^2_B}}}\right),
\end{align}
and redefined the function
\begin{equation}
\tilde{g}(k_i) = |\omega| g(k_i),
\end{equation}
\end{widetext}
and $\theta(x)=1$ for $x\geq 0$ and $\theta(x)=0$ for $x< 0$.

\subsection{Magneto-optical conductivity}
The conductivity tensor is obtained from the response function as
\begin{equation} \label{eq:mocon}
    \sigma_{\alpha\beta}(\omega)=\frac{1}{i\omega}\left[\chi_{\alpha \beta}(\omega)-\chi_{\alpha \beta}(0)\right].
\end{equation}
It follows that the response functions vanish when $\omega=0$, since $E_n(k_z)\neq E_{n+1}(k_z)$ for any $n$ (see \hyperref[app:I]{Appendix \ref{app:I}} for details). Moreover, for $\omega \neq 0$, the corresponding conductivity components read
\begin{align}\label{eq:sigmaxx}
   &\sigma_{xx}(\omega) = \frac{1}{\omega}\left\{\text{Im}\left[\chi_{xx}(\omega)\right]-i \ \text{Re}\left[\chi_{xx}(\omega)\right]\right\}\\
    \nonumber
    &=\frac{1}{\omega}\left\{\text{Im}\left[\chi_{xx}(\omega)\right]-i \text{PV} \int_{-\infty}^{+\infty}\frac{d\omega'}{\pi}\frac{\text{Im}\left[\chi_{xx}(\omega')\right]}{\omega'-\omega}\right\} ,
    \end{align}
  and  
    \begin{align}\label{eq:sigmaxy}
    &\sigma_{xy}(\omega) = \frac{1}{\omega}\left\{\text{Im}\left[\chi_{xy}(\omega)\right]-i \ \text{Re}\left[\chi_{xy}(\omega)\right]\right\}\\
    \nonumber
    &=-\frac{1}{\omega}\left\{ \ \text{PV}\int_{-\infty}^{+\infty}\frac{d\omega'}{\pi}\frac{\text{Re}\left[\chi_{xy}(\omega')\right]}{\omega'-\omega}+i\text{Re}\left[\chi_{xy}(\omega)\right]\right\}.
\end{align}

We will focus on the $\sigma_{xx}(\omega)$-component for the results. 

\section{Results}\label{sec:III}
Previous studies of magneto-optical conductivity mainly discuss systems expanded to linear order in momentum \cite{AC2013,TCG2016,UB2016,YYY2016}. However, in the type-II regime this results in systems with infinite Fermi pockets, which necessarily requires an unphysical cutoff to yield a finite magneto-optical conductivity. Here, we show how to partially overcome these shortcomings and apply it to two experimentally relevant situations---one where we expand the momentum to cubic order along the direction of the tilt to generate a type-II Weyl system with over-tilted electron and hole pockets, and one system additionally hosting trivial Fermi pockets away from the Weyl node. Thanks to the higher-order terms in momentum, the Fermi pockets are closed \cite{TBUK2017,TBK2018}, removing the need of a cutoff in momentum.

\subsection{Cubic Dispersion}
Let us consider a Weyl-like Hamiltonian where we take cubic terms into account in the direction of a potential tilting. This is necessary to get finite sized Fermi pockets in the type-II-regime as pointed out in Refs. \cite{TBUK2017,TBK2018}, allowing us to perform the calculations without the need of introducing a nonphysical large cutoff to regulate the integral in Eq. (\ref{eq:response function}). The Hamiltonian for this model takes the particular form
\begin{equation} \label{eq:cubham}
    H(\bk)=\hbar v_F\left[\left(-\eta k_z + \gamma k_z^3\right)\sigma_0 + \left(k_z+\beta k_z^3\right)\sigma_z+k_x\sigma_x+k_y\sigma_y\right],
\end{equation}
where $\beta$ and $\gamma$ are the cubic order corrections. The parameter $\eta$ is exactly the tilt of the Weyl cone. Noting that the Fermi velocity of the Weyl semimetal TaAs is $0.286$ eV$\cdot${\AA} \cite{Xuetal2016}, we will use the exemplary, yet realistic, value $\hbar v_F = 0.3$ eV$\cdot${\AA} in what follows, and it will be understood in the equations below, to avoid cumbersome notation. When applying a magnetic field along the $k_z$-direction, and introducing the creation and annihilation operators, Eq. (\ref{eq:cubham}) becomes
\begin{equation}
    H(\bk)=\begin{pmatrix}\left(1-\eta\right)k_z + \left(\gamma+\beta\right)k_z^3 & \frac{\sqrt{2}}{l_B}a^{\dagger}\\\frac{\sqrt{2}}{l_B}a & -\left(1+\eta\right)k_z+\left(\gamma-\beta\right)k_z^3 \end{pmatrix}.
\end{equation}
The eigenvalues of $H$, corresponding to the Landau levels of the systems read
\begin{equation}
    E_{n,\lambda}(k_z) = \left[-\eta k_z+\gamma k_z^3 +\lambda\sqrt{\left(k_z+\beta k_z^3\right)^2 + \frac{2n}{l^2_B}}\right],
\end{equation}
when $n\neq 0$ and $\lambda=\pm 1$, and the chiral level is given by
\begin{equation}
E_0(k_z) = \left[(1-\eta)k_z+(\beta+\gamma)k_z^3\right].
\end{equation}
We note however that a different Fermi velocity would yield the exact same result, if the parameters of the model would be modified accordingly. For example, if $ v_F \to \xi v_F$, for some constant $\xi$, then by changing $(1-\eta)\to \xi^{-1}(1-\eta)$, $(\gamma+\beta)\to \xi^{-1}(\gamma+\beta)$ and $B\to \xi^{-2}B$, the two models would yield exactly the same result. Further details are treated in \hyperref[sec:appc]{Appendix \ref{sec:appc}}.
 
The eigenstates schematically attain the same form as in Eqs. \eqref{eq:es1}, \eqref{eq:es2} and \eqref{eq:es3}, with $g(k_z) = k_z+\beta k_z^3$. Since $g(k_z)$ is a cubic polynomial, the magneto-optical conductivity can be evaluated completely analytically. The contribution will come from all the real values of $k_z$ satisfying an equation involving $g(k_z)$. The aforementioned equation is analytically solvable since $g(k_z)$ is a cubic polynomial. Further details on the form of this particular equation can be found in.

Next, we will present our results for three representative cases: (1) an untilted system, (2) a tilted type-I system, and (3) a tilted type-II system, separately.

\subsubsection{Untilted System}
\begin{figure}[H]
 \hbox to \linewidth{ \hss
\includegraphics[width=\linewidth]{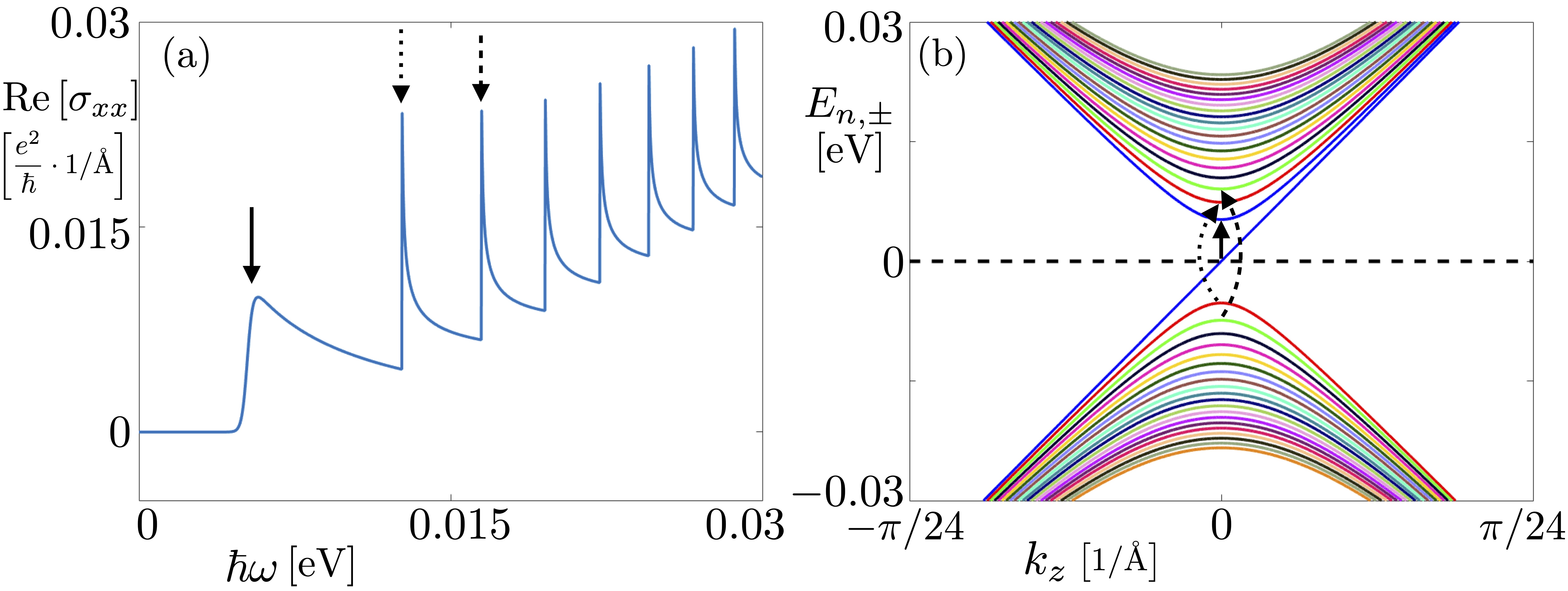}
 \hss}
\caption{Real part of the diagonal component of the magneto-optical conductivity for an untilted Weyl semimetal (a) with dispersion cubic in $k_z$ (b). Here, we have used $\eta=\gamma=0$, $\beta=\frac{1}{6}$, $B=10$ T, $\mu = 0$, and $T=5$ K. The peak corresponding to the transition from $n$ to $n+1$ occurs, as expected, at $\hbar \omega = \sqrt{\frac{2(n+1)}{l_B^2}}+\sqrt{\frac{2n}{l_B^2}}$. The small bump corresponds exactly to the transition from the chiral level to the first Landau level, appearing at $\hbar \omega = \sqrt{\frac{2}{l_B^2}} = 0.017$ eV. The arrows indicate the first few transitions.}
\label{fig:untiltedsigmaxx}
\end{figure}

For a Weyl semimetal with untilted dispersion $h(k_z)=0$, we essentially expect to reproduce the result that is known from previous literature, since there are no Fermi pockets to be closed. In Fig. \ref{fig:untiltedsigmaxx}(a), we illustrate the real part of $\sigma_{xx}$ as a function of $\hbar\omega$ along with the corresponding Landau levels in Fig. \ref{fig:untiltedsigmaxx}(b), and we note that our work agrees with, e.g., Refs. \cite{AC2013,TCG2016}. In particular,  asymmetric peaks with a sharp onset appear at $\hbar \omega = \sqrt{\frac{2n}{l_B^2}}+\sqrt{\frac{2(n+1)}{l_B^2}}$. The $n^{\text{th}}$ peak corresponds to the inter-band transition from $n$ to $n+1$ illustrated in Fig. \ref{fig:untiltedsigmaxx}(b), and they appear because of the selection rules coming from the matrix elements in the response function. Note, the smaller rounded initial hump corresponds to the transition from the chiral level to the first Landau level.

\subsubsection{Tilted Type-I System}
\begin{figure}[H]
 \hbox to \linewidth{ \hss
\includegraphics[width=\linewidth]{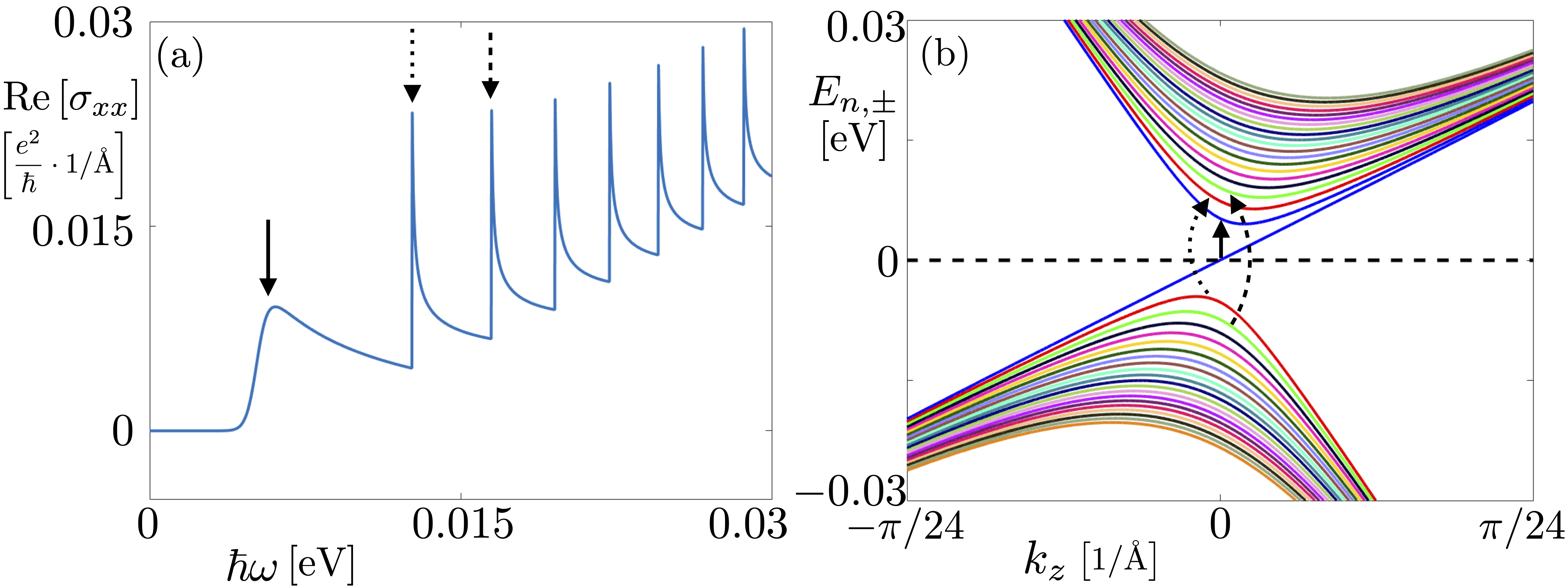}
 \hss}
\caption{Real part of the diagonal component of the magneto-optical conductivity for a type-I Weyl semimetal (a) with dispersion cubic in $k_z$ (b). Here, we have used $\eta=\frac{1}{2}$, $\gamma = \frac{1}{10}$ $\beta=\frac{1}{6}$, $B=10$ T, $\mu = 0$, and $T=5$ K. Both the main peaks and the initial bump appear at the same positions as in the untilted case. Again, the arrows indicates the first few transitions.}
\label{fig:type1sigmaxx}
\end{figure}

For a type-I Weyl semimetal, the main features of the real part of $\sigma_{xx}$ are similar to those of the untilted case, which can be seen in Fig. $\ref{fig:type1sigmaxx}$(a). This is because as long as the cone is not overtilted, there will be no Fermi pockets, and thus no additional transitions are expected. The sharp peaks appear at the same values of $\hbar\omega$, and we understand them in the same way. Also, the initial bump, corresponding to the transition from the chiral level to the first Landau level, appear at the expected position. 

\subsubsection{Tilted Type-II System}
\begin{figure}[H]
 \hbox to \linewidth{ \hss
\includegraphics[width=\linewidth]{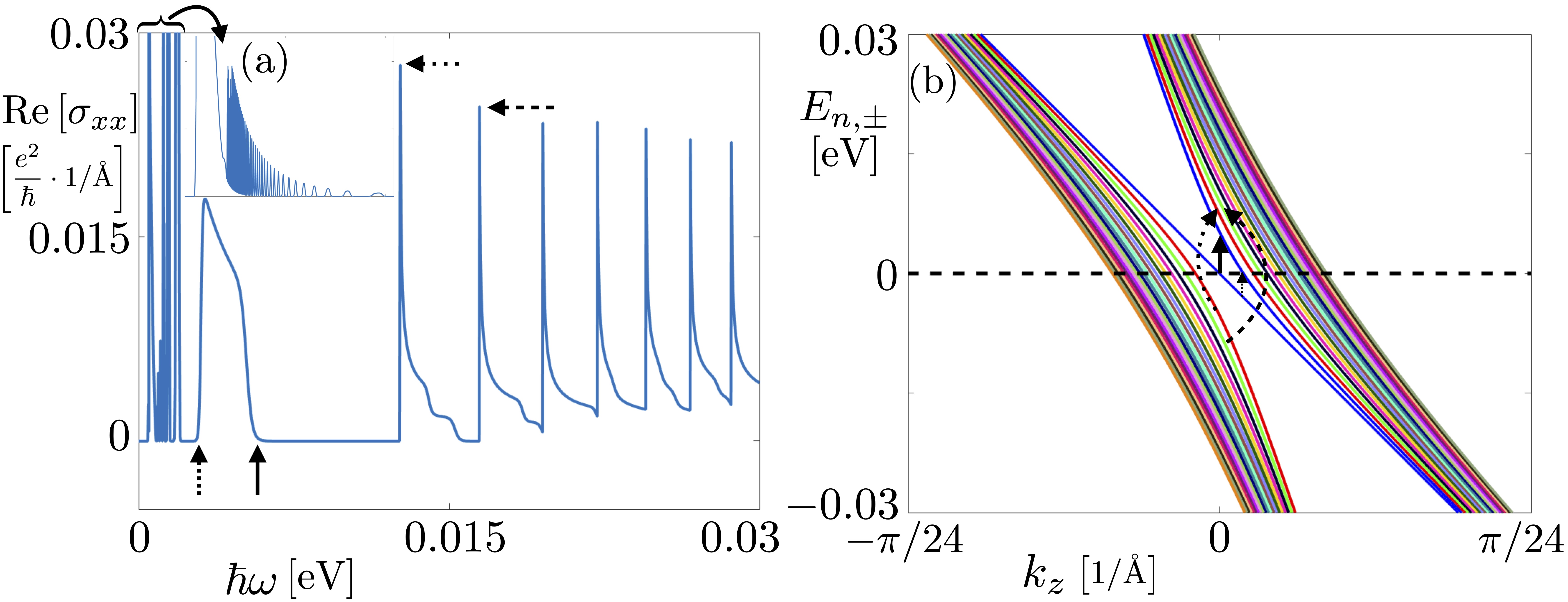}
 \hss}
\caption{Real part of the diagonal component of the magneto-optical conductivity for a type-II Weyl semimetal (a) with dispersion cubic in $k_z$ (b). Here, we have used $\eta=2$, $\gamma=\frac{1}{10}$, $\beta=\frac{1}{6}$, $B=10$ T, $\mu = 0$, and $T=5$ K. Similarly to the untilted and type-I cases, the main peaks occur at $\hbar \omega = \sqrt{\frac{2(n+1)}{l_B^2}}+\sqrt{\frac{2n}{l_B^2}}$, but additional bumps are appearing in between the main peaks. Also, intraband transitions occur for small $\hbar \omega$, which is displayed in the top left corner. Furthermore, the initial bump is now coming from two allowed transitions between the chiral Level and the first Landau level, one appearing at $k_z=0$ and one at nonzero $k_z$, resulting in the sudden decay which is not present in the previous cases.}
\label{fig:type2sigmaxx}
\end{figure}

When the Weyl cone is over-tilted, i.e. tipped over such that Fermi pockets are formed, the magneto-optical conductivity changes qualitatively, see Fig. \ref{fig:type2sigmaxx}(a). The sharp onset of the asymmetric peaks are still appearing at $\hbar \omega = \sqrt{\frac{2n}{l^2_B}}+\sqrt{\frac{2(n+1)}{l^2_B}}$, but they are not decaying in the same way because of the asymmetry between positive/negative momentum transitions, see Fig. \ref{fig:type2sigmaxx}(b). These features are also noticed in Refs. \onlinecite{TCG2016,UB2016,YYY2016}. 

Furthermore, additional first broad peak originating from transitions involving the chiral level has a different form, which is a direct consequence of the existence of Fermi pockets. The lowest and highest energy transitions between the chiral level and the first Landau level are indicated by arrows in Fig. \ref{fig:type2sigmaxx}(b). 

Finally, new excitations appear at lowest frequency which are a direct consequence of the Fermi pockets. At nonzero $k_z$, transitions between Landau levels of the same chirality are now possible. These intraband transitions happen exactly when, e.g., the $n$th Landau level of the electron/hole pocket crosses the Fermi energy and the $(n+1)$th Landau level does not.

\subsection{Trivial Fermi Pockets}
Finally, we address another interesting and experimentally relevant example where the tools developed in this work can be applied to systems hosting trivial Fermi pockets, e.g. we can model systems with additional trivial Fermi pockets which are not in direct connection to the Weyl point. Most Weyl semimetal materials indeed feature such additional trivial pockets, and with this motivation we investigate their effects on the magneto-optical conductivity.  

\begin{figure}[H]
 \hbox to \linewidth{ \hss
\includegraphics[width=\linewidth]{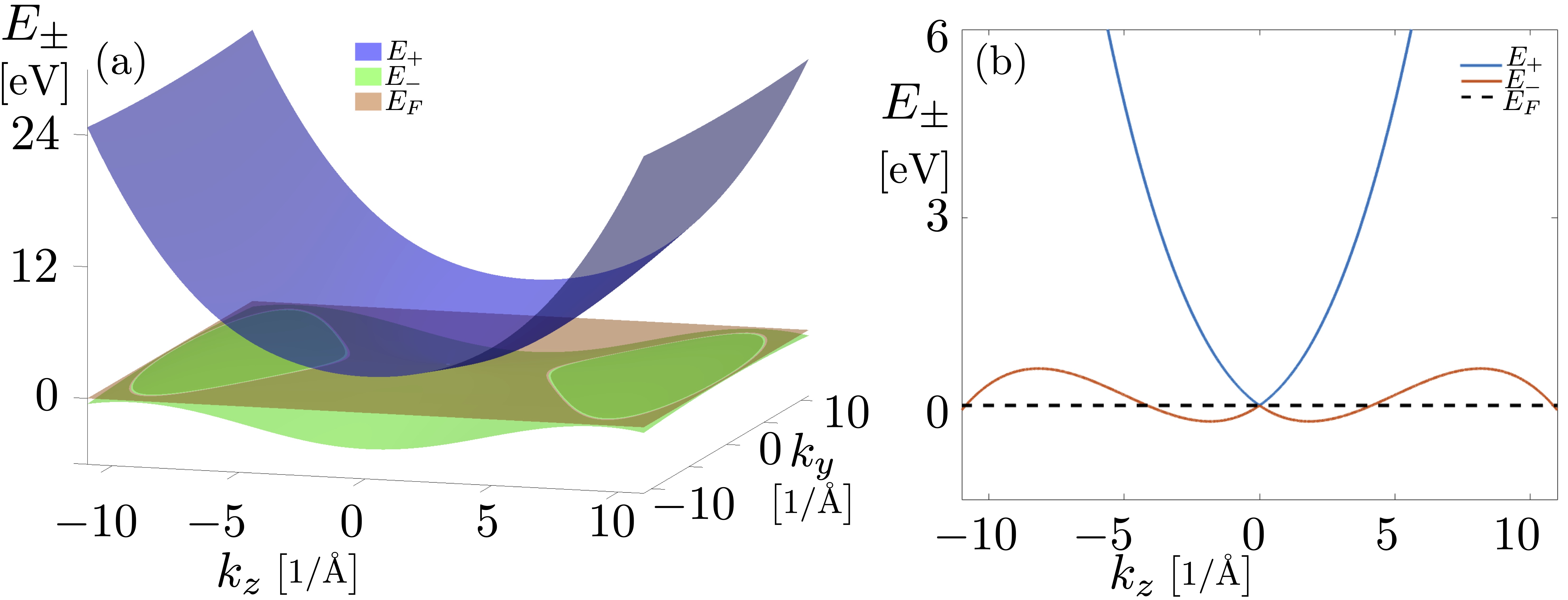}
 \hss}
\caption{Band structure of the system described by the Hamiltonian in Eq. (\ref{eq:hamfermi}) with $\beta = \frac{1}{45}$, $\alpha = \frac{1}{3}$, and $\hbar v_F = 1$ eV$\cdot$\AA. The lower energy band is forming trivial Fermi pockets away from the Weyl node, which are shown to be closed in all directions in momentum space.}
\label{fig:fermipocketenergeis}
\end{figure}

To achieve the desired band structure, we assume a Hamiltonian on the form
\begin{equation} \label{eq:hamfermi}
    H(\bk) = \hbar v_F\left[\sigma_x k_x + \sigma_y k_y+(k_z+\beta k_z^3)\sigma_z + \alpha k_z^2 \sigma_0\right],
\end{equation}
and assign $\beta = \frac{1}{45}$ and $\alpha = \frac{1}{3}$. Again, we use $\hbar v_F = 0.3$ eV$\cdot$ \AA, and remove it from the following equations. The band structure of this system is displayed in Fig. \ref{fig:fermipocketenergeis}. The lower band now crosses the Fermi energy not only at the Weyl point but also away from it, and thus forming (closed) trivial Fermi pockets.

Next we apply the magnetic field in the $k_z$-direction and the energy levels become,
\begin{equation}
    E_{n,\lambda}(k_z) = \alpha k_z^2 +\lambda\sqrt{\left(k_z+\beta k_z^3\right)^2 + \frac{2n}{l_B^2}},
\end{equation}
and
\begin{equation}
E_0(k_z) = k_z+\alpha k_z^2+\beta k_z^3.
\end{equation}
Note, that the difference in energy between different Landau levels is scaling with square root of the magnetic field around the Weyl node. However, because of our construction of the trivial pockets via a higher order expansion, the splitting in energy around the trivial pockets is much smaller, since it scales with a higher power.

\begin{figure}[H]
 \hbox to \linewidth{ \hss
\includegraphics[width=\linewidth]{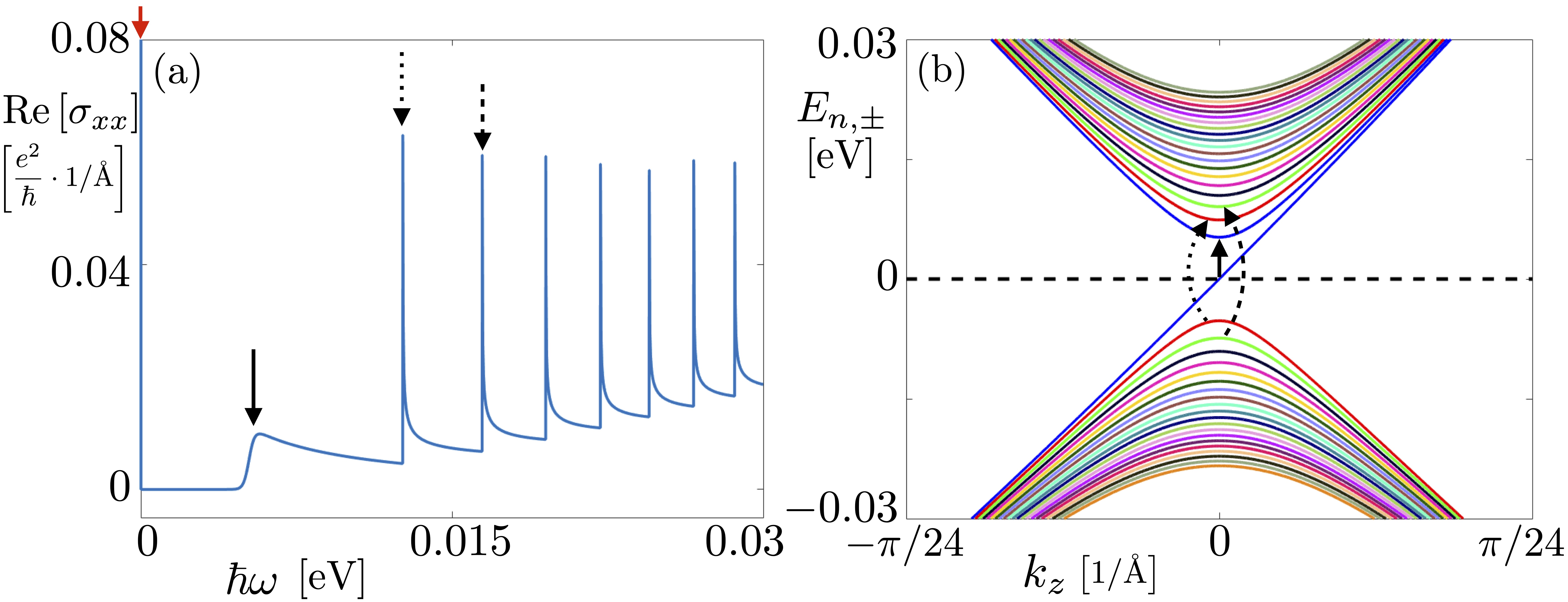}
 \hss}
\caption{Real part of the diagonal component of the magneto-optical conductivity for an untilted Weyl system hosting trivial Fermi pockets (a), along with the corresponding Landau levels (b). Here, we have used $\alpha = \frac{1}{3}$, $\beta = \frac{1}{45}$, $B=10$ T, $\mu = 0$, and $T=5$ K. One would expect the trivial Fermi pockets to contribute with additional transitions for small $\hbar \omega$, but for the chosen value of the magnetic field, these are not visible since they are expected to appear for extremely small values of $\hbar \omega$. The trivial pockets are furthermore not changing the behavior of the magneto-optical conductivity on a global scale.}
\label{fig:fermipocketsigmaxx}
\end{figure}

Using Eqs. \eqref{eq:resfunxx} and \eqref{eq:mocon}, the magneto-optical conductivity can be directly calculated, and $\text{Re}\left[\sigma_{xx}(\omega)\right]$ is shown in Fig. \ref{fig:fermipocketsigmaxx}(a) together with a plot of the Landau levels zoomed in around the Weyl point in Fig. \ref{fig:fermipocketsigmaxx}(b). Despite having a fundamentally different band structure, the behavior of the magneto-optical conductivity on a global scale is not changed---the main peaks appear at the same values of $\hbar \omega$ because we are probing transitions with zero momentum transfer.

\begin{figure}[H]
 \hbox to \linewidth{ \hss
\includegraphics[width=\linewidth]{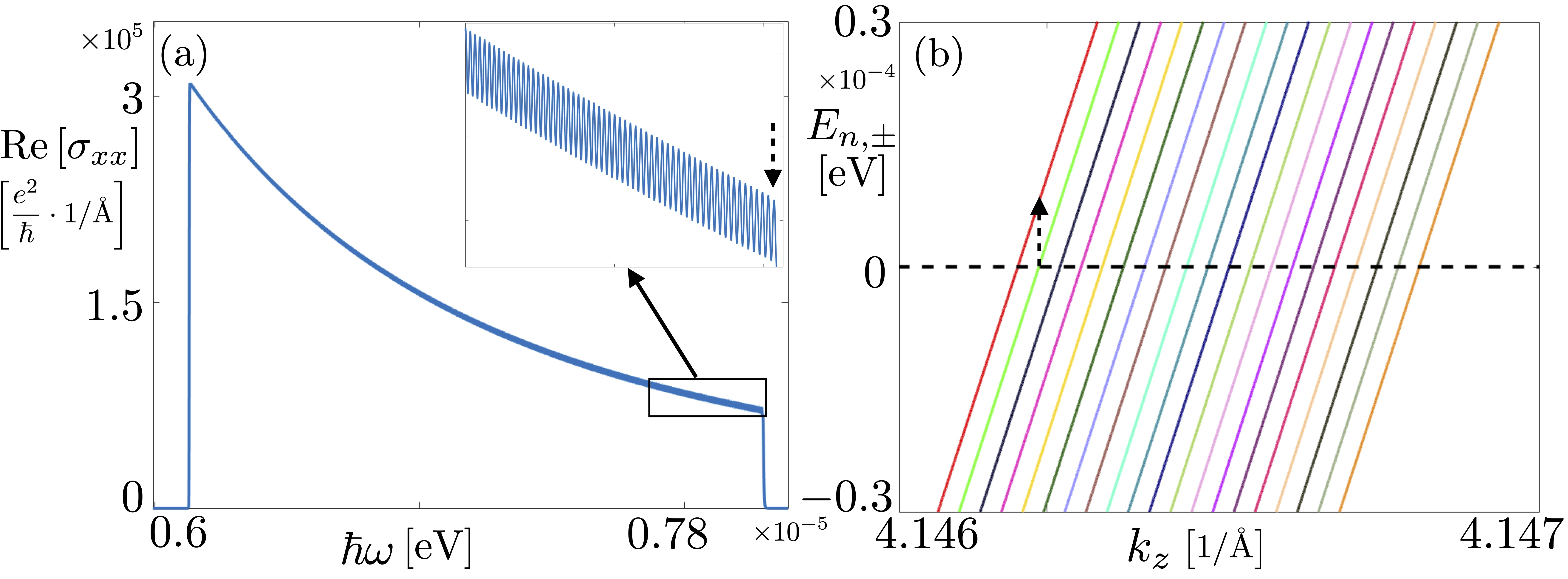}
 \hss}
\caption{Real part of the diagonal component of the magneto-optical conductivity at small $\hbar \omega$ for an untilted Weyl system hosting trivial Fermi pockets (a), along with the corresponding Landau levels (b) at the border of one of the Fermi pockets. Here, we have used $\alpha = \frac{1}{3}$, $\beta = \frac{1}{45}$, $B=10$ T, $\mu = 0$, and $T=5$ K. In these systems, as in type-II Weyl semimetals, intraband transitions occur. Here they happen at energy scales $10^{-5}$ eV. The small Landau level spacing give rise to an almost oscillatory behavior with the transitions appearing very close to each other.}
\label{fig:fermipocketsigmaxxhighB}
\end{figure}

However, there appears an additional sharp spike close to $\hbar \omega = 0$ in Fig. \ref{fig:fermipocketsigmaxx}, marked by the red arrow. A zoom into these low-energy transitions is presented in Fig. \ref{fig:fermipocketsigmaxxhighB}(a), along with the band structure at one end of one of the Fermi pockets [Fig. \ref{fig:fermipocketsigmaxxhighB}(b)]. These additional intraband transitions happen at nonzero $k_z$ due to the existence of trivial Fermi pockets, which are not in direct connection with a Weyl point.  Interestingly, the amplitude of these transitions is much larger than that of the conventional interband transitions from the Weyl physics. 

Within our model the high intensity contribution from the trivial pockets is always separated from the Weyl contribution appearing at much lower frequency. However, this is an artefact of our specific model and the higher order expansion of only the $k_z$ dispersion. In real materials the trivial pockets will come with their own cyclotron frequency such that their contribution generically overlaps with the Weyl node transitions. Additionally, scattering by disorder is expected to smear out the sharp features from the Landau level transitions such that the main contribution is a smooth background in the magneto-optical conductivity, see Fig. \ref{fig:fermipocketsigmaxxhighB}(a). Due to the much larger magnitude of the intraband transitions, it is thus likely that they will mask the conventional Weyl peaks in materials, and could therefore provide a simple explanation for why these peaks have not been observed. 

\section{Discussion and outlook}\label{sec:IV}
We have developed a general analytical theory for computing the magneto-optical conductivity for Weyl semimetals with arbitrary tilt in a magnetic field. We applied our tools to evaluate the magneto-optics in different representative classes of  Weyl systems, and the inclusion of higher-order momentum terms has allowed us to go beyond the linearized regime studied previously. In particular, we treated type-II systems with closed  Fermi pockets. Furthermore our treatment allowed us to investigate the effect of additional trivial Fermi pockets away from the Weyl node as is often the case for material candidates. 

We verified that the higher order terms in general do not lead to drastic changes of the magneto-optical conductivity contribution from the interband Weyl node transitions. However, additional low frequency peaks appear from intraband transitions due to  the presence of Fermi pockets. While the latter always appear at very low energies for pockets from over-tilted type-II Weyl systems, this is generally not the case for additional trivial Fermi pockets away from the Weyl node. Their contribution is of intra-band character with an enhanced intensity which is argued to mask the contribution of the Weyl node, thus providing a simple explanation for their absence in experiments. 

The method described in this work can also be used to evaluate the magneto-optical conductivity for other systems of the form in Eq. \eqref{eq:genham}. Thus, one could in principle scan through different models of interest, possibly going beyond linear response to study interaction effects, i.e. for collective modes. This means that the general expression derived in this work can be used to evaluate the magneto-optical conductivity for models describing actual materials. In this way, it would be possible to prove or disprove our hypothesis regarding why the peak structure from the interband transitions visible in theoretical models are not observed in the corresponding experiments. This would, however, require the relevant band structures to be derived from first principles, such as DFT. This goes beyond the scope of this work, and is therefore left as a future, interesting study to be carried out.  Alternatively, a study of the transverse conductivity could be extended to layered systems for investigating possible signatures of a three dimensional quantum Hall effect. 
Finally, our basic recipe to go beyond the linearized Weyl regime, which allows us to model closed Fermi pockets, should be applied to experimental observables probing nonzero momentum transitions like dynamical spin and charge susceptibilities where qualitative changes to the simple linearized models are expected.


\begin{acknowledgments}
We thank M. Udagawa and M. Trescher for fruitful discussions and related collaborations. E.J.B. and M.S. are supported by the Swedish Research Council (VR) and the Wallenberg Academy Fellows program of the Knut and Alice Wallenberg Foundation. J.K. acknowledges helpful discussions with A. Pronin. 
\end{acknowledgments}

\appendix


\begin{widetext}
	
	\section{Detailed calculations}\label{app:I}
	
	As anticipated, the optical conductivity is calculated, in the one-loop approximation, using linear response theory. The linear response function is given by
	\begin{align}
	\chi_{\alpha\beta}(\omega) &:= \frac{1}{2\pi l^2_B}\sum_{n,n'}\sum_{\lambda,\lambda '}\int \frac{dk_z}{2\pi}\frac{f\left[E_{n,\lambda}(k_z)\right]-f\left[E_{n',\lambda '}(k_z)\right]}{\hbar\omega+E_{n,\lambda}(k_z)-E_{n',\lambda '}(k_z)+i\epsilon} \nonumber
	\\
	&\quad \times \langle{\psi_{n,\lambda}(k_z)}|j_{\alpha}|{\psi_{n',\lambda '}(k_z)}\rangle\langle{\psi_{n',\lambda'}(k_z)}|j_{\beta}|{\psi_{n,\lambda }(k_z)}\rangle,
	\end{align}
	where $f(E)=\frac{1}{1+e^{\frac{E-\mu}{k_B T}}}$ is the Fermi-Dirac distribution function, $T$ is the temperature and $j_{\alpha}$ is a current operator, given by 
	\begin{equation}
	j_{\alpha}=\frac{e}{\hbar}\frac{\partial H}{\partial \Pi_{\alpha}}.
	\end{equation}
	From this, the conductivity tensor is obtained by
	\begin{equation}
	\sigma_{\alpha\beta}(\omega)=\frac{1}{i\omega}\left[\chi_{\alpha \beta}(\omega)-\chi_{\alpha \beta}(0)\right].
	\end{equation}
	We are interested in computing the transverse components, namely $\sigma_{xx}$ and $\sigma_{xy}$, for which the relevant current operators become
	\begin{equation}
	j_x=ev_F\sigma_x, \quad j_y=-ev_F\sigma_y.
	\end{equation}
	
	Let us first derive a closed-form expression for the response functions for general functions $g(k_z)$ and $h(k_z)$. We will treat $\chi_{xx}(\omega)$ and $\chi_{xy}(\omega)$ in parallel. The respective matrix elements take the following form
	\begin{align}
	&\quad\langle{\psi_{n,\lambda}(k_z)}|j_{x}|{\psi_{n',\lambda '}(k_z)}\rangle\langle{\psi_{n',\lambda'}(k_z)}|j_{x}|{\psi_{n,\lambda }(k_z)}\rangle
	\nonumber
	\\
	&=e^2v_F^2\left\{\left[u_{n,\lambda}(k_z)\right]^2\left[v_{n',\lambda'}(k_z)\right]^2\delta_{n+1,n'} + \left[v_{n,\lambda}(k_z)\right]^2\left[u_{n',\lambda'}(k_z)\right]^2\delta_{n-1,n'}\right\},
	\end{align}
	
	\begin{align}
	&\quad\langle{\psi_{n,\lambda}(k_z)}|j_{x}|{\psi_{n',\lambda '}(k_z)}\rangle\langle{\psi_{n',\lambda'}(k_z)}|j_{x}|{\psi_{n,\lambda }(k_z)}\rangle
	\nonumber
	\\
	&=ie^2v_F^2\left\{\left[u_{n,\lambda}(k_z)\right]^2\left[v_{n',\lambda'}(k_z)\right]^2\delta_{n+1,n'} - \left[v_{n,\lambda}(k_z)\right]^2\left[u_{n',\lambda'}(k_z)\right]^2\delta_{n-1,n'}\right\}.
	\end{align}
	Thus, the corresponding response functions read
	\begin{align}
	\chi_{xx}(\omega) &=\frac{ e^2v_F^2}{8\pi l^2_B}\sum_{n,\lambda,\lambda'}\int \frac{dk_z}{2\pi}\left(\left\{f\left[E_{n,\lambda}(k_z)\right]-f\left[E_{n+1,\lambda'}(k_z)\right]\right\}\frac{\left[1+\frac{g(k_z)}{\lambda\sqrt{g^2(k_z)+\frac{2n}{l_B^2}}}\right]\left[1-\frac{g(k_z)}{\lambda'\sqrt{g^2(k_z)+\frac{2(n+1)}{l_B^2}}}\right]}{\hbar\omega-\left[E_{n,\lambda}(k_z)-E_{n+1,\lambda'}(k_z)\right]+i\epsilon} \right. \nonumber
	\\
	&\left.-\left\{f\left[E_{n,\lambda'}(k_z)\right]-f\left[E_{n+1,\lambda}(k_z)\right]\right\}\frac{\left[1-\frac{g(k_z)}{\lambda\sqrt{g^2(k_z)+\frac{2(n+1)}{l_B^2}}}\right]\left[1+\frac{g(k_z)}{\lambda'\sqrt{g^2(k_z)+\frac{2n}{l_B^2}}}\right]}{\hbar\omega-\left[E_{n+1,\lambda}(k_z)-E_{n,\lambda'}(k_z)\right]+i\epsilon}\right),
	\\
	\chi_{xy}(\omega) &=\frac{i e^2v_F^2}{8\pi l^2_B}\sum_{n,\lambda,\lambda'}\int \frac{dk_z}{2\pi}\left(\left\{f\left[E_{n,\lambda}(k_z)\right]-f\left[E_{n+1,\lambda'}(k_z)\right]\right\}\frac{\left[1+\frac{g(k_z)}{\lambda\sqrt{g^2(k_z)+\frac{2n}{l_B^2}}}\right]\left[1-\frac{g(k_z)}{\lambda'\sqrt{g^2(k_z)+\frac{2(n+1)}{l_B^2}}}\right]}{\hbar\omega-\left[E_{n,\lambda}(k_z)-E_{n+1,\lambda'}(k_z)\right]+i\epsilon} \right. \nonumber
	\\
	&\left.+\left\{f\left[E_{n,\lambda'}(k_z)\right]-f\left[E_{n+1,\lambda}(k_z)\right]\right\}\frac{\left[1-\frac{g(k_z)}{\lambda\sqrt{g^2(k_z)+\frac{2(n+1)}{l_B^2}}}\right]\left[1+\frac{g(k_z)}{\lambda'\sqrt{g^2(k_z)+\frac{2n}{l_B^2}}}\right]}{\hbar\omega-\left[E_{n+1,\lambda}(k_z)-E_{n,\lambda'}(k_z)\right]+i\epsilon}\right).
	\end{align}
	
	At this point, we invoke the Kramers-Kroning relations which show that the real and imaginary parts of the response function are not completely independent pieces of information but rather, each part is totally determined by the other. Thus, the information in the response function is duplicated and it is enough to compute either the real or the imaginary part. For the sake of simplicity of the calculations, we choose to compute the imaginary part of $\chi_{xx}$ and the real part of $\chi_{xy}$, respectively. 
	
	Considering the clean limit $\epsilon\to 0$ and making use of the functional identity 
	\begin{equation}
	\frac{1}{x\pm i\epsilon}=\text{PV}\left(\frac{1}{x}\right)\mp i\pi\delta(x),
	\end{equation}
	this yields
	\begin{align}
	\text{Im}\left[\chi_{xx}(\omega)\right]&=-\frac{e^2v_F^2}{8 l^2_B}\sum_{n,\lambda,\lambda'}\int \frac{dk_z}{2\pi}\left(\left\{f\left[E_{n,\lambda}(k_z)\right]-f\left[E_{n+1,\lambda'}(k_z)\right]\right\} \right. \nonumber
	\\
	&\quad \times \left[1+\frac{g(k_z)}{\lambda\sqrt{g^2(k_z)+\frac{2n}{l_B^2}}}\right]\left[1-\frac{g(k_z)}{\lambda'\sqrt{g^2(k_z)+\frac{2(n+1)}{l_B^2}}}\right]\delta\left\{\hbar\omega-\left[ E_{n,\lambda}(k_z)-E_{n+1,\lambda'}(k_z)\right]\right\} \nonumber
	\\
	&\quad-\left\{f\left[E_{n,\lambda'}(k_z)\right]-f\left[E_{n+1,\lambda}(k_z)\right]\right\} \nonumber
	\\
	&\quad \times\left.\left[1-\frac{g(k_z)}{\lambda\sqrt{g^2(k_z)+\frac{2(n+1)}{l_B^2}}}\right]\left[1+\frac{g(k_z)}{\lambda'\sqrt{g^2(k_z)+\frac{2n}{l_B^2}}}\right]\delta\left\{\hbar\omega-\left[E_{n+1,\lambda}(k_z)-E_{n,\lambda'}(k_z)\right]\right\}\right),
	\\
	\text{Re}\left[\chi_{xy}(\omega)\right]&=-\frac{e^2v_F^2}{8 l^2_B}\sum_{n,\lambda,\lambda'}\int \frac{dk_z}{2\pi}\left(\left\{f\left[E_{n,\lambda}(k_z)\right]-f\left[E_{n+1,\lambda'}(k_z)\right]\right\} \right. \nonumber
	\\
	&\quad \times \left[1+\frac{g(k_z)}{\lambda\sqrt{g^2(k_z)+\frac{2n}{l_B^2}}}\right]\left[1-\frac{g(k_z)}{\lambda'\sqrt{g^2(k_z)+\frac{2(n+1)}{l_B^2}}}\right]\delta\left\{\hbar\omega-\left[ E_{n,\lambda}(k_z)-E_{n+1,\lambda'}(k_z)\right]\right\} \nonumber
	\\
	&\quad+\left\{f\left[E_{n,\lambda'}(k_z)\right]-f\left[E_{n+1,\lambda}(k_z)\right]\right\} \nonumber
	\\
	&\quad \times \left.\left[1-\frac{g(k_z)}{\lambda\sqrt{g^2(k_z)+\frac{2(n+1)}{l_B^2}}}\right]\left[1+\frac{g(k_z)}{\lambda'\sqrt{g^2(k_z)+\frac{2n}{l_B^2}}}\right]\delta\left\{\hbar\omega-\left[E_{n+1,\lambda}(k_z)-E_{n,\lambda'}(k_z)\right]\right\}\right).
	\end{align}
	When performing the sum over $\lambda$ and $\lambda'$, we will in total end up with eight Dirac $\delta$-distributions
	\begin{align}
	\delta\left\{ \hbar\omega-\left[E_{n,+}(k_z)-E_{n+1,+}(k_z)\right]\right\} &= \delta\left\{ \hbar\omega-\left[E_{n+1,-}(k_z)-E_{n,-}(k_z)\right]\right\},\label{eq:delta1}
	\\
	\delta\left\{ \hbar\omega-\left[E_{n,-}(k_z)-E_{n+1,-}(k_z)\right]\right\} &= \delta\left\{ \hbar\omega-\left[E_{n+1,+}(k_z)-E_{n,+}(k_z)\right]\right\},\label{eq:delta2}
	\\
	\delta\left\{ \hbar\omega-\left[E_{n,+}(k_z)-E_{n+1,-}(k_z)\right]\right\} &= \delta\left\{ \hbar\omega-\left[E_{n+1,+}(k_z)-E_{n,-}(k_z)\right]\right\},\label{eq:delta3}
	\\
	\delta\left\{ \hbar\omega-\left[E_{n,-}(k_z)-E_{n+1,+}(k_z)\right]\right\} &= \delta\left\{\hbar\omega-\left[E_{n+1,-}(k_z)-E_{n,+}(k_z)\right]\right\},\label{eq:delta4}
	\end{align}
	which tell us that the only values of $k_z$ contributing to the integral are those satisfying the $\delta$-distribution. These values of $k_z$ are obtained by solving for when the arguments of the above $\delta$-distributions vanish. Specifically, we have to solve the following
	\begin{align}
	\hbar \omega -\hbar v_F\left[ \sqrt{g^2(k_z)+\frac{2n}{l^2_B}}-\sqrt{g^2(k_z)+\frac{2(n+1)}{l^2_B}}\right] &= 0,
	\\
	\hbar \omega -\hbar v_F\left[ -\sqrt{g^2(k_z)+\frac{2n}{l^2_B}}+\sqrt{g^2(k_z)+\frac{2(n+1)}{l^2_B}}\right] &= 0,
	\\
	\hbar \omega -\hbar v_F\left[ \sqrt{g^2(k_z)+\frac{2n}{l^2_B}}+\sqrt{g^2(k_z)+\frac{2(n+1)}{l^2_B}}\right] &= 0,
	\\
	\hbar \omega -\hbar v_F\left[ -\sqrt{g^2(k_z)+\frac{2n}{l^2_B}}-\sqrt{g^2(k_z)+\frac{2(n+1)}{l^2_B}}\right] &= 0,
	\end{align}
	which in the end boils down to solving
	\begin{equation} \label{eq:zerosg}
	g(k_z)=\pm\sqrt{\frac{ v_F^2}{\omega^2l^4_B}-\frac{(2n+1)}{l^2_B}+\frac{\omega^2}{4v_F^2}}
	\end{equation}
	for $k_z$. We note that $g(k_z)$ has to be real-valued, which gives us a maximum allowed Landau level
	\begin{equation}
	n_{\text{max}} = \left\lfloor \frac{\left(2 v_F^2-\omega^2l^2_B \right)^2}{8v_F^2\omega^2l^2_B}\right\rfloor.
	\end{equation}
	
	However, we must further note that these solutions are only true for some specific values of $\omega$. In fact, each distribution will result in different constraints on $\omega$. While Eq. \eqref{eq:delta1} will be valid for $\omega<0$ and $\omega^2<\frac{2v_F^2}{l^2_B}$, Eq. \eqref{eq:delta2} is valid for $\omega>0$ and $\omega^2<\frac{2v_F^2}{l^2_B}$, Eq. \eqref{eq:delta3} is valid for $\omega<0$ and $\omega^2>\frac{2v_F^2}{l^2_B}$ and finally, Eq. \eqref{eq:delta4} is valid for $\omega>0$ and $\omega^2>\frac{2v_F^2}{l^2_B}$. To not loose track, the detailed calculations leading to the constraints on $\omega$ can be found in \hyperref[app:II]{Appendix \ref{app:II}}. Taking this into account, we proceed to perform the integration over $k_z$. To do this, we recall the property of the Dirac $\delta$ distribution
	\begin{equation}
	\int dx \ f(x) \ \delta\left[g(x)\right]=\sum_{x_i}\frac{f(x_i)}{\mid g'(x_i) \mid },
	\end{equation}
	where ${x_i}$ are the zeros of $g(x)$, i.e. $g(x_i)=0$. In order to simplify notation, let us introduce the following dimensionless coefficients:
	\begin{align}
	A^{\pm,\pm}_n(k_i) &:= f\left[E_{n,\pm}(k_i)\right]-f\left[E_{n+1,\pm}(k_i)\right]=\frac{\sinh{\frac{E_{n+1,\pm}(k_i)-E_{n,\pm}(k_i)}{2k_B T}}}{\cosh{\frac{E_{n+1,\pm}(k_i)+E_{n,\pm}(k_i)-2\mu}{2k_B T}}+\cosh{\frac{E_{n+1,\pm}(k_i)-E_{n,\pm}(k_i)}{2k_B T}}},
	\\
	B^{\pm}_n(k_i) &:= 1\pm\frac{g(k_i)}{\sqrt{g^2(k_i)+\frac{2n}{l^2_B}}},
	\\
	C^{\pm,\pm}_{n,m}(k_i) &:= \left| \frac{\hbar v_F\sqrt{g^2(k_i)+\frac{2n}{l^2_B}}\sqrt{g^2(k_i)+\frac{2m}{l^2_B}}}{g(k_i)g'(k_i)\left[E_{n,\pm}(k_i)-E_{m,\pm}(k_i)\right]}\right|,
	\end{align}
	where $k_i$ are the values of $k_z$ satisfying Eq. \ref{eq:zerosg}. In terms of these, the response functions explicitly read
	\begin{align}
	&\text{Im}\left[\chi_{xx}(\omega)\right]=-\frac{e^2v_F^2}{16\pi l^2_B}\sum_{n=0}^{n_{\text{max}}}\sum_{i=1}^{2m} \nonumber
	\\
	&\left\{ A_n^{++}(k_i)\left[B^+_n(k_i)B^-_{n+1}(k_i)C_{n,n+1}^{++}(k_i)\theta(-\omega)\theta(\frac{2v_F^2}{l^2_B}-\omega^2)-B_n^+(k_i)B_{n+1}^-(k_i)C^{++}_{n+1,n}(k_i)\theta(\omega)\theta(\frac{2v_F^2}{l^2_B}-\omega^2)\right]\right.\nonumber
	\\
	&+A^{--}_n(k_i)\left[B_n^-(k_i)B^+_{n+1}(k_i)C^{--}_{n,n+1}(k_i)\theta(\omega)\theta(\frac{2v_F^2}{l^2_B}-\omega^2)-B_n^-(k_i)B^+_{n+1}(k_i)C^{--}_{n+1,n}(k_i)\theta(-\omega)\theta(\frac{2v_F^2}{l^2_B}-\omega^2)\right] \nonumber
	\\
	&+A^{+-}_n(k_i)\left[B_n^+(k_i)B^+_{n+1}(k_i)C^{+-}_{n,n+1}(k_i)\theta(\omega)\theta(\omega^2-\frac{2v_F^2}{l^2_B})-B_n^+(k_i)B^+_{n+1}(k_i)C^{-+}_{n+1,n}(k_i)\theta(-\omega)\theta(\omega^2-\frac{2v_F^2}{l^2_B})\right]\nonumber
	\\
	&\left.+A^{-+}_n(k_i)\left[B_n^-(k_i)B^-_{n+1}(k_i)C^{-+}_{n,n+1}(k_i)\theta(-\omega)\theta(\omega^2-\frac{2v_F^2}{l^2_B})-B_n^-(k_i)B^-_{n+1}(k_i)C^{+-}_{n+1,n}(k_i)\theta(\omega)\theta(\omega^2-\frac{2v_F^2}{l^2_B})\right] \right\},
	\end{align}
	\begin{align}
	&\text{Re}\left[\chi_{xy}(\omega)\right]=-\frac{e^2v_F^2}{16\pi l^2_B}\sum_{n=0}^{n_{\text{max}}}\sum_{i=1}^{2m} \nonumber
	\\
	&\left\{ A_n^{++}(k_i)\left[B^+_n(k_i)B^-_{n+1}(k_i)C_{n,n+1}^{++}(k_i)\theta(-\omega)\theta(\frac{2v_F^2}{l^2_B}-\omega^2)+B_n^+(k_i)B_{n+1}^-(k_i)C^{++}_{n+1,n}(k_i)\theta(\omega)\theta(\frac{2v_F^2}{l^2_B}-\omega^2)\right]\right.\nonumber
	\\
	&+A^{--}_n(k_i)\left[B_n^-(k_i)B^+_{n+1}(k_i)C^{--}_{n,n+1}(k_i)\theta(\omega)\theta(\frac{2v_F^2}{l^2_B}-\omega^2)+B_n^-(k_i)B^+_{n+1}(k_i)C^{--}_{n+1,n}(k_i)\theta(-\omega)\theta(\frac{2v_F^2}{l^2_B}-\omega^2)\right] \nonumber
	\\
	&+A^{+-}_n(k_i)\left[B_n^+(k_i)B^+_{n+1}(k_i)C^{+-}_{n,n+1}(k_i)\theta(\omega)\theta(\omega^2-\frac{2v_F^2}{l^2_B})+B_n^+(k_i)B^+_{n+1}(k_i)C^{-+}_{n+1,n}(k_i)\theta(-\omega)\theta(\omega^2-\frac{2v_F^2}{l^2_B})\right]\nonumber
	\\
	&\left.+A^{-+}_n(k_i)\left[B_n^-(k_i)B^-_{n+1}(k_i)C^{-+}_{n,n+1}(k_i)\theta(-\omega)\theta(\omega^2-\frac{2v_F^2}{l^2_B})+B_n^-(k_i)B^-_{n+1}(k_i)C^{+-}_{n+1,n}(k_i)\theta(\omega)\theta(\omega^2-\frac{2v_F^2}{l^2_B})\right] \right\}.
	\end{align}
	
	Note that the factors $C^{\pm,\pm}_{n,n+1}(k_i)$ will in fact be the same in all the terms due to the absolute value in their definition and the accompanying Heaviside $\theta$ functions and thus take the form
	\begin{equation}
	C_{n,n+1}^{\pm,\pm}=C_{n,n+1}=C_{n+1,n}=\abs{\frac{\left(\frac{\omega^2}{4v_{F}^{2}}-\frac{v_{F}^{2}}{\omega^2l_{B}^4}\right)}{g(k_i)g'(k_i)\frac{\omega}{v_F}}}.
	\end{equation}
	Using this, we arrive at the final expressions 
	\begin{align}
	\text{Im}\left[\chi_{xx}(\omega)\right]&=\frac{e^2v_F^2}{16\pi l^2_B}\sum_{n=0}^{n_{\text{max}}}\sum_{i=1}^{2m}\left(\left[D^+_n(k_i,\omega,l_B)+D^-_n(k_i,\omega,l_B)\right]C_{n,n+1}(k_i)\right. 
	\nonumber
	\\
	&\quad \times \left.\left\{\left[F^{-+}_{n,n+1}(k_i)+F^{+-}_{n,n+1}(k_i)\right]\theta(\frac{2v_F^2}{l^2_B}-\omega^2)+\left[F^{++}_{n,n+1}(k_i)+F^{--}_{n,n+1}(k_i)\right]\theta(\omega^2-\frac{2v_F^2}{l^2_B})\right\}\right),
	\end{align}
	\begin{align}
	\text{Re}\left[\chi_{xy}(\omega)\right]&= -\frac{e^2v_F^2}{16\pi l^2_B}\sum_{n=0}^{n_{\text{max}}}\sum_{i=1}^{2m}\left(\left[D^+_n(k_i,\omega,l_B)-D^-_n(k_i,\omega,l_B)\right]C_{n,n+1}(k_i) \right.
	\nonumber
	\\
	&\quad \times \left.\left\{\left[F^{-+}_{n,n+1}(k_i)+F^{+-}_{n,n+1}(k_i)\right]\theta(\frac{2v_F^2}{l^2_B}-\omega^2)+\left[F^{++}_{n,n+1}(k_i)+F^{--}_{n,n+1}(k_i)\right]\theta(\omega^2-\frac{2v_F^2}{l^2_B})\right\}\right), 
	\end{align}
	where
	\begin{align}
	D^{\pm} &= \frac{\sinh{\frac{\hbar\omega}{2k_BT}}}{\cosh{\left[\frac{2\hbar v_Fh(k_i)\pm\frac{2\hbar^2v_{F}^{2}}{\hbar\omega l^2_B}-2\mu}{2k_BT}\right]}+\cosh\frac{\hbar\omega}{2k_BT}},
	\\
	F^{\pm\pm}_{n,m} &= \left[1\pm \frac{\tilde{g}(k_i)}{\omega\sqrt{g^2(k_i)+\frac{2n}{l^2_B}}}\right]\left[1\pm\frac{\tilde{g}(k_i)}{\omega\sqrt{g^2(k_i)+\frac{2m}{l^2_B}}}\right],
	\\
	\tilde{g}(k_i) &= \abs{\omega} g(k_i).
	\end{align}
	
	One thing we should note is that the response functions vanish when $\omega=0$, since $E_n(k_z)\neq E_{n+1}(k_z)$ for any $n$. Thus, the corresponding conductivity components for $\omega \neq 0$ associated to $\text{Im}\left[\chi_{xx}(\omega)\right]$ and $\text{Re}\left[\chi_{xy}(\omega)\right]$ read
	\begin{align}
	\text{Re}\left[\sigma_{xx}(\omega)\right] &= \frac{\text{Im}\left[\chi_{xx}(\omega)\right]}{\omega},
	\\
	\text{Im}\left[\sigma_{xy}(\omega)\right] &=-\  \frac{\text{Re}\left[\chi_{xy}(\omega)\right]}{\omega}.
	\end{align}
	\section{Dirac deltas and frequency bounds}\label{app:II}
	The different $\delta$-distributions we obtained in order to perform the integration over $k_z$ are actually leading to additional constraints, which are displayed in terms of the Heaviside functions appearing with every term in the final expressions. Here, we will explicitly derive the constraints on $\omega$ in terms of $l_B$. The fact that some terms are only giving a contribution for positive and negative $\omega$ respectively is obvious, and will therefore not be dealt with. The different $\delta$-distributions are (\ref{eq:delta1}--\ref{eq:delta4})
	which will be treated case by case.
	\subsection{Case 1}
	The equations to solve are
	\begin{align}
	\hbar\omega-\left[E_{n+1,+}(k_i)-E_{n,+}(k_i)\right]&=0, \label{eq:delta1eq1}
	\\
	\hbar\omega-\left[E_{n,-}(k_i)-E_{n+1,-}(k_i)\right]&=0,\label{eq:delta1eq2}
	\end{align}
	which both require $\omega>0$, and they both explicitly become
	\begin{equation}
	\hbar\omega - \hbar v_F\left(\sqrt{\frac{v_F^2}{\omega^2l_B^4}-\frac{2n+1}{l^2_B}+\frac{\omega^2}{4v_F^2}+\frac{2n+1}{l^2_B}}-\sqrt{ \frac{v_F^2}{\omega^2l_B^4}-\frac{2n+1}{l^2_B}+\frac{\omega^2}{4v_F^2}+\frac{2n}{l^2_B}}\right)=0.
	\end{equation}
	Simplifying this yields
	\begin{equation}
	\hbar\omega-\hbar v_F\left[\frac{1}{2|\omega|v_F}\left(\frac{2v_F^2}{l^2_B}+\omega^2\right) -\frac{1}{2|\omega|v_F}\left|\frac{2v_F^2}{l^2_B}-\omega^2\right|\right]=0.
	\end{equation}
	Using that $\omega>0$, and first assuming that $\omega^2<\frac{2v_F^2}{l^2_B}$ we get
	\begin{equation}
	\hbar\omega-\frac{2\hbar\omega^2}{2\omega}=0,
	\end{equation}
	which is always satisfied, so for $\omega^2<\frac{2v_F^2}{l^2_B}$ the equation can be solved, and thus the $\delta$ distribution is giving a contribution. Let us now consider the case when $\omega^2>\frac{2v_F^2}{l^2_B}$, then we get
	\begin{equation}
	\hbar\omega-\hbar v_F\left[\frac{1}{2\omega v_F}\left(\frac{2v_F^2}{l^2_B}+\omega^2\right)+\frac{1}{2\omega v_F}\left(\frac{2v_F^2}{l^2_B}-\omega^2\right)\right]=0,
	\end{equation}
	which is equivalent to
	\begin{equation}
	\omega^2=\frac{2v_F^2}{l^2_B}.
	\end{equation}
	But this is a contradiction since the initial assumption for this case was that $\omega^2>\frac{2v_F^2}{l^2_B}$ strictly. Therefore, we see that Eqs. \eqref{eq:delta1eq1} and  \eqref{eq:delta1eq2}, corresponding to the $\delta$ distributions Eq. \eqref{eq:delta1}, are only solvable when $\omega>0$ and furthermore $\omega^2<\frac{2v_F^2}{l^2_B}$.
	
	\subsection{Case 2}
	The equations to solve are
	\begin{align}
	\hbar\omega-\left[E_{n+1,-}(k_i)-E_{n,-}(k_i)\right]&=0, \label{eq:delta2eq1}
	\\
	\hbar\omega-\left[E_{n,+}(k_i)-E_{n+1,+}(k_i)\right]&=0,\label{eq:delta2eq2}
	\end{align}
	which both require $\omega<0$, and they both explicitly become
	\begin{equation}
	\hbar\omega - \hbar v_F\left(-\sqrt{\frac{v_F^2}{\omega^2l_B^4}-\frac{2n+1}{l^2_B}+\frac{\omega^2}{4v_F^2}+\frac{2(n+1)}{l^2_B}}+\sqrt{ \frac{v_F^2}{\omega^2l_B^4}-\frac{2n+1}{l^2_B}+\frac{\omega^2}{4v_F^2}+\frac{2n}{l^2_B}}\right)=0.
	\end{equation}
	Simplifying this yields
	\begin{equation}
	\hbar\omega-\hbar v_F\left[-\frac{1}{2|\omega|v_F}\left(\frac{2v_F^2}{l^2_B}+\omega^2\right) +\frac{1}{2|\omega|v_F}\left|\frac{2v_F^2}{l^2_B}-\omega^2\right|\right]=0.
	\end{equation}
	Using that $\omega<0$, and first assuming that $\omega^2<\frac{2v_F^2}{l^2_B}$ we get
	\begin{equation}
	\hbar\omega-\frac{2\hbar\omega^2}{2\omega}=0,
	\end{equation}
	which is always satisfied, so for $\omega^2<\frac{2v_F^2}{l^2_B}$ the equation can be solved, and thus the $\delta$ distribution is giving a contribution. Let us now consider the case when $\omega^2>\frac{2v_F^2}{l^2_B}$, then we get
	\begin{equation}
	\hbar\omega-\hbar v_F\left[\frac{1}{2\omega v_F}\left(\frac{2v_F^2}{l^2_B}+\omega^2\right)+\frac{1}{2\omega v_F}\left(\frac{2v_F^2}{l^2_B}-\omega^2\right)\right]=0,
	\end{equation}
	which is equivalent to
	\begin{equation}
	\omega^2=\frac{2v_F^2}{l^2_B}.
	\end{equation}
	But this is again a contradiction since the initial assumption for this case was that $\omega^2>\frac{2v_F^2}{l^2_B}$ strictly. Therefore, we see that Eqs. \eqref{eq:delta2eq1} and \eqref{eq:delta2eq2}, corresponding to the $\delta$ distributions Eq. \eqref{eq:delta2}, are only solvable when $\omega<0$ and furthermore $\omega^2<\frac{2v_F^2}{l^2_B}$.
	
	\subsection{Case 3}
	The equations to solve are
	\begin{align}
	\hbar\omega-\left[E_{n+1,+}(k_i)-E_{n,-}(k_i)\right]&=0, \label{eq:delta3eq1}
	\\
	\hbar\omega-\left[E_{n,+}(k_i)-E_{n+1,-}(k_i)\right]&=0,\label{eq:delta3eq2}
	\end{align}
	which both require $\omega>0$, and they both explicitly become
	\begin{equation}
	\hbar\omega - \hbar v_F\left(\sqrt{\frac{v_F^2}{\omega^2l_B^4}-\frac{2n+1}{l^2_B}+\frac{\omega^2}{4v_F^2}+\frac{2(n+1)}{l^2_B}}-\sqrt{ \frac{v_F^2}{\omega^2l_B^4}-\frac{2n+1}{l^2_B}+\frac{\omega^2}{4v_F^2}+\frac{2n}{l^2_B}}\right)=0.
	\end{equation}
	Simplifying this yields
	\begin{equation}
	\hbar\omega-\hbar v_F\left[\frac{1}{2|\omega|v_F}\left(\frac{2v_F^2}{l^2_B}+\omega^2\right) +\frac{1}{2|\omega|v_F}\left|\frac{2v_F^2}{l^2_B}\right|\right]=0.
	\end{equation}
	Using that $\omega>0$, and first assuming that $\omega^2>\frac{2v_F^2}{l^2_B}$ we get
	\begin{equation}
	\hbar\omega-\frac{2\hbar\omega^2}{2\omega}=0,
	\end{equation}
	which is always satisfied, so for $\omega^2>\frac{2v_F^2}{l^2_B}$ the equation can be solved, and thus the $\delta$ distribution is giving a contribution. Let us now consider the case when $\omega^2<\frac{2v_F^2}{l^2_B}$, then we get
	\begin{equation}
	\hbar\omega-\left[\frac{1}{2\omega v_F}\left(\frac{2v_F^2}{l^2_B}+\omega^2\right)+\frac{1}{2\omega v_F}\left(\frac{2v_F^2}{l^2_B}-\omega^2\right)\right]=0,
	\end{equation}
	which is equivalent to
	\begin{equation}
	\omega^2=\frac{2v_F^2}{l^2_B}.
	\end{equation}
	But again this is a contradiction since the initial assumption for this case was that $\omega^2<\frac{2v_F^2}{l^2_B}$ strictly. Therefore, we see that Eqs. \eqref{eq:delta3eq1} and \eqref{eq:delta3eq2}, corresponding to the $\delta$ distributions Eq. \eqref{eq:delta3}, are only solvable when $\omega>0$ and furthermore $\omega^2>\frac{2v_F^2}{l^2_B}$.
	
	\subsection{Case 4}
	The equations to solve are
	\begin{align}
	\hbar\omega-\left[E_{n+1,-}(k_i)-E_{n,+}(k_i)\right]&=0, \label{eq:delta4eq1}
	\\
	\hbar\omega-\left[E_{n,-}(k_i)-E_{n+1,+}(k_i)\right]&=0,\label{eq:delta4eq2}
	\end{align}
	which both require $\omega<0$, and they both explicitly become
	\begin{equation}
	\hbar\omega - \hbar v_F\left(\sqrt{\frac{v_F^2}{\omega^2l_B^4}-\frac{2n+1}{l^2_B}+\frac{\omega^2}{4v_F^2}+\frac{2(n+1)}{l^2_B}}-\sqrt{ \frac{v_F^2}{\omega^2l_B^4}-\frac{2n+1}{l^2_B}+\frac{\omega^2}{4v_F^2}+\frac{2n}{l^2_B}}\right)=0.
	\end{equation}
	Simplifying this yields
	\begin{equation}
	\hbar\omega-\hbar v_F\left[-\frac{1}{2|\omega|v_F}\left(\frac{2v_F^2}{l^2_B}+\omega^2\right) -\frac{1}{2|\omega|v_F}\left|\frac{2v_F^2}{l^2_B}-\omega^2\right|\right]=0.
	\end{equation}
	Using that $\omega<0$, and first assuming that $\omega^2>\frac{2v_F^2}{l^2_B}$ we get
	\begin{equation}
	\hbar\omega-\frac{2\hbar\omega^2}{2\omega}=0,
	\end{equation}
	which is always satisfied, so for $\omega^2>\frac{2v_F^2}{l^2_B}$ the equation can be solved, and thus the $\delta$ distribution is giving a contribution. Let us now consider the case when $\omega^2<\frac{2v_F^2}{l^2_B}$, then we get
	\begin{equation}
	\hbar\omega-\hbar v_F\left[\frac{v_F^2}{2\omega}\left(\frac{2v_F^2}{l^2_B}+\omega^2\right)+\frac{1}{2\omega v_F}\left(\frac{2v_F^2}{l^2_B}-\omega^2\right)\right]=0,
	\end{equation}
	which is equivalent to
	\begin{equation}
	\omega^2=\frac{2v_F^2}{l^2_B}.
	\end{equation}
	But this is a contradiction, since the initial assumption for this case was that $\omega^2<\frac{2v_F^2}{l^2_B}$ strictly. Therefore, we see that Eqs. \eqref{eq:delta4eq1} and  \eqref{eq:delta4eq2}, corresponding to the $\delta$ distributions Eq. \eqref{eq:delta4}, are only solvable when $\omega<0$ and furthermore $\omega^2>\frac{2v_F^2}{l^2_B}$.
	
	\section{The Fermi Velocity and Re-Scaling Properties}\label{sec:appc}
	In this section, we elaborate on the role played by the Fermi velocity and how one can readily rescale the results of a given calculation to obtain results for other parameter values. Consider two different Hamiltonians,
	\begin{align}
	H&=\hbar v_F\begin{pmatrix}h(k_z)+g(k_z) & \frac{\sqrt{2}}{l_B}a^{\dagger}\\\frac{\sqrt{2}}{l_B}a & h(k_z)-g(k_z) \end{pmatrix},
	\\
	\tilde{H}&=\hbar \tilde{v}_F\begin{pmatrix}\tilde{h}(k_z)+\tilde{g}(k_z) & \frac{\sqrt{2}}{\tilde{l}_B}a^{\dagger}\\\frac{\sqrt{2}}{\tilde{l}_B}a & \tilde{h}(k_z)-\tilde{g}(k_z) \end{pmatrix},
	\end{align}
	with corresponding eigensystem,
	\begin{align}
	E_{n,\lambda}(k_z,l_B) &= \hbar v_F\left[ h(k_z)+\lambda\sqrt{g^2(k_z)+\frac{2n}{l^2_B}}\right]; \quad E_{0}(k_z) = \hbar v_F\left[h(k_z)+g(k_z)\right],
	\\
	\tilde{E}_{n,\lambda}(k_z,\tilde{l}_B) &= \hbar \tilde{v}_F\left[ \tilde{h}(k_z)+\lambda\sqrt{\tilde{g}^2(k_z)+\frac{2n}{\tilde{l}^2_B}}\right]; \quad \tilde{E}_{0}(k_z) = \hbar \tilde{v}_F\left[\tilde{h}(k_z)+\tilde{g}(k_z)\right],
	\\
    u_{n,\lambda}(k_z) &= \sqrt{\frac{1}{2}\left[1+\frac{g(k_z)}{\lambda \sqrt{g^2(k_z)+\frac{2n}{l_B^2}}}\right]}; \quad v_{n,\lambda}(k_z) = \sqrt{\frac{1}{2}\left[1-\frac{g(k_z)}{\lambda \sqrt{g^2(k_z)+\frac{2n}{l_B^2}}}\right]},
    \\
    \tilde{u}_{n,\lambda}(k_z) &= \sqrt{\frac{1}{2}\left[1+\frac{\tilde{g}(k_z)}{\lambda \sqrt{\tilde{g}^2(k_z)+\frac{2n}{\tilde{l}_B^2}}}\right]}; \quad \tilde{v}_{n,\lambda}(k_z) = \sqrt{\frac{1}{2}\left[1-\frac{\tilde{g}(k_z)}{\lambda \sqrt{\tilde{g}^2(k_z)+\frac{2n}{\tilde{l}_B^2}}}\right]},
    \\
    \psi_{n,\lambda}(k_z)&=\begin{pmatrix}\lambda u_{n,\lambda}(k_z) \\ v_{n,\lambda}(k_z)\end{pmatrix}; \quad \tilde{\psi}_{n,\lambda}(k_z)=\begin{pmatrix}\lambda \tilde{u}_{n,\lambda}(k_z) \\ \tilde{v}_{n,\lambda}(k_z)\end{pmatrix}; \quad \psi_0(k_z) = \tilde{\psi}_0(k_z) = \begin{pmatrix} 1\\ 0\end{pmatrix}
	\end{align}
	Take $\tilde{v}_F = \xi v_F$ for some real number $\xi$. Then we would like to investigate when $H$ and $\tilde{H}$ provide the same physics. The first requirement would be that the eigenvalues are equal, which corresponds to,
	\begin{equation}
	 h(k_z)+\lambda\sqrt{g^2(k_z)+\frac{2n}{l^2_B}} = \xi\left[ \tilde{h}(k_z)+\lambda\sqrt{\tilde{g}^2(k_z)+\frac{2n}{\tilde{l}^2_B}}\right]
	\end{equation}
	which is clearly true if and only if,
	\begin{equation} \label{eq:constraint}
	h(k_z) = \xi \tilde{h}(k_z); \quad g(k_z) = \xi \tilde{g}(k_z); \quad B = \xi^2 \tilde{B}
	\end{equation}
	Clearly, $E_0(k_z) = \tilde{E}_0(k_z)$ holds true if Eq. \eqref{eq:constraint} is satisfied. We furthermore note that $\psi_{n,\lambda}(k_z) = \tilde{\psi}_{n,\lambda}(k_z)$ when imposing the constraint in Eq. \eqref{eq:constraint}. This then means that different values of the Fermi velocity can be simultaneously studied, if one allows for a rescaling of $g(k_z)$, $h(k_z)$ and the magnetic field $B$. Hence, it may seem slightly redundant to plug realistic values of the Fermi velocity if the values of the constant parameters in $g(k_z)$ and $h(k_z)$ are not known individually---what matters physically is not the overall factor, but rather the values of all the remaining free parameters. Therefore, to actually try to compute the magneto-optical conductivity for a realistic material, the full structure of the valence and the conduction band has to be known.
	\end{widetext}

\end{document}